\documentclass[aps,prd,10pt,superscriptaddress,amsfonts,amssymb,amsmath,preprintnumbers,tightenlines,floats,notitlepage,letterpaper,twocolumn,nofootinbib,longbibliography]{revtex4-2}

\usepackage{slashed}
\usepackage{amsmath}
\usepackage{amssymb}
\usepackage{amsthm}
\usepackage{hyperref}
\usepackage{indentfirst}
\usepackage{psfrag}
\usepackage{cleveref}
\usepackage[utf8]{inputenc}
\usepackage{graphicx}
\usepackage{aas_macros}

\hypersetup{colorlinks=true, linkcolor=blue, urlcolor=blue, citecolor=blue, linktocpage=true}

\newcommand{\be}{\begin{equation}}
\newcommand{\ee}{\end{equation}}
\newcommand{\ba}{\begin{align*}}
\newcommand{\ea}{\end{align*}}
\newcommand{\beqa}{\begin{eqnarray}}
\newcommand{\eeqa}{\end{eqnarray}}
\newcommand{\bseq}{\begin{subequations}}
\newcommand{\eseq}{\end{subequations}}
\newcommand*\diff{\mathrm{d}}

\let\limitint\int 
\renewcommand{\int}{\limitint \!} 

\newcommand{\D}{\mathcal{D}}

\renewcommand\d{\partial}
\newcommand\G{\Gamma}
\newcommand\e{\text{e}}
\renewcommand\l{\lambda}
\newcommand\vf{\varphi}
\renewcommand\a{\alpha}

\renewcommand\Im{\text{Im}}
\newcommand\const{\text{const}}

\renewcommand\o{\omega}

\newcommand{\sh}{\mathop{\rm sinh}\nolimits}
\newcommand{\ch}{\mathop{\rm cosh}\nolimits}

\begin{document}

\title{Thermal False Vacuum Decay Is More Than It Seems}





\author{Dalila P\^irvu}
\email{dalila.pirvu@uwaterloo.ca}
\affiliation{\perimeter}
\affiliation{\waterloo}

\author{Andrey Shkerin}
\email{ashkerin@perimeterinstitute.ca}
\thanks{(corresponding author)}
\affiliation{\perimeter}

\author{Sergey Sibiryakov}
\email{ssibiryakov@perimeterinstitute.ca}
\affiliation{\perimeter}
\affiliation{\mcmaster}

\newcommand{\perimeter}{Perimeter Institute for Theoretical Physics, 31 Caroline St N, Waterloo, ON N2L 2Y5, Canada}
\newcommand{\waterloo}{Department of Physics \& Astronomy, University of Waterloo, Waterloo, ON N2L 3G1, Canada}
\newcommand{\mcmaster}{Department of Physics \& Astronomy, McMaster University, 1280 Main Street West, Hamilton, ON L8S 4M1, Canada}

\begin{abstract}

We study the decay of a thermally excited metastable vacuum in classical field theory using real-time numerical simulations. We find a significantly lower decay rate than predicted by standard thermal theory at moderate temperatures, $E_b/T\sim 10$, where $E_b$ is the critical bubble energy. The discrepancy is due to the violation of thermal equilibrium during the critical bubble nucleation and is reduced if thermalization is enhanced by introduction of dissipation and thermal noise. 
We formulate a condition for the system to remain in equilibrium during the nucleation process and show that it is generally violated in weakly coupled field theories. Nevertheless, we argue that the violation of thermal equilibrium becomes irrelevant for the false vacuum decay rate at sufficiently low temperatures and the standard thermal rate is recovered.

\end{abstract}

\maketitle

\section{Introduction}
\label{sec:intro}

The decay of a metastable state (false vacuum) plays an important role in many branches of physics. It corresponds to first-order phase transitions in condensed matter systems and relativistic field theories \cite{Onuki_2002, Quiros:1999jp}. In cosmology, such phase transitions have been extensively studied in the context of baryon asymmetry generation \cite{Bodeker:2020ghk} and as possible sources of gravitational waves \cite{Caprini:2018mtu, Caprini:2019egz}. The current electroweak vacuum of the Standard Model may be metastable \cite{Buttazzo:2013uya, Bednyakov:2015sca}, implying its decay in the future. There are several proposals to realize false vacuum decay using cold atom systems \cite{Fialko:2014xba, Fialko:2016ggg, Braden:2017add, Billam:2018pvp, Braden:2019vsw, Billam:2021nbc, Jenkins:2023eez, Jenkins:2023npg}, and the first successful experiment was reported in \cite{Zenesini:2023afv,Cominotti:2025qia}.

In many physical situations, the initial state of the system is an equilibrium thermal state around the false vacuum with some temperature $T$. The traditional approach to this case is based on the Euclidean path integral method \cite{Kobzarev:1974cp, Coleman:1977py, Callan:1977pt, Linde:1980tt, Linde:1981zj}, which relates the decay rate to the imaginary part of the metastable vacuum free energy. At high enough temperatures, the transition proceeds via formation of a critical bubble -- an unstable solution of the classical field equations that can decay both to the false and the true vacuum. It corresponds to the saddle point of the potential barrier separating the two vacua. The Euclidean approach then yields the decay rate in the form \cite{Affleck:1980ac},\footnote{We use the system of units $c=\hbar=k_B=1$ and define the rate as the probability of decay per unit time and volume.}
\be \label{GammaAffleck}
    \Gamma=\frac{\omega_-}{\pi T}\cdot\frac{\Im F}{\cal V}\, ,    
\ee
where $\omega_-$ is the growth rate of the critical bubble's unstable mode and ${\cal V}$ is the volume of the system. The imaginary part of the free energy in the false vacuum contains the Boltzmann suppression by the critical bubble energy, $\Im F\propto \e^{-E_b/T}$, as well as the determinant of the operator describing small fluctuations around it \cite{Weinberg:2012pjx}.

At $\omega_-\ll T\ll E_b$, the result (\ref{GammaAffleck}) can also be obtained by purely classical methods. Langer \cite{Langer:1969bc} considered a classical multi-dimensional statistical system with dissipation and noise provided by an \emph{external} heat bath and controlled by the friction parameter $\eta$. False vacuum decay then occurs as a result of diffusion in phase space, and the solution of the corresponding Fokker-Planck equation yields the rate \cite{Hanggi:1990zz},
\be \label{GammaLanger}
    \Gamma=\frac{1}{\pi T}\left(\sqrt{\omega_-^2+\frac{\eta^2}{4}}-\frac{\eta}{2}\right) \cdot\frac{\Im F}{\cal V}\, ,
\ee
which reduces to (\ref{GammaAffleck}) in the limit $\eta\to 0$.

The Euclidean approach can tell us little about the dynamics of bubble nucleation. Instead, this can be captured by real-time numerical simulations \cite{Grigoriev:1988bd, Grigoriev:1989je, Grigoriev:1989ub, Ambjorn:1990wn, Valls:1990, Alford:1991qg, Alford:1993zf, Alford:1993ph, Borsanyi:2000ua, Braden:2018tky, Batini:2023zpi}. These have revealed rich phenomena, including oscillon precursors and non-zero bubble velocities \cite{Aguirre:2011ac, Gleiser:1991rf, Gleiser:2004iy, Gleiser:2007ts, Pirvu:2023plk}.
In this work, we continue the real-time study of thermal false vacuum decay, focusing on the accurate measurement of its rate, including both the Boltzmann suppression and the fluctuation determinant.
Surprisingly, we find deviations from Eqs.~(\ref{GammaAffleck}), (\ref{GammaLanger}) at moderate temperatures $E_b/T\sim 10$ accessible to direct simulations.
The deviations are traced back to a breakdown of thermal equilibrium during bubble nucleation.
We formulate a 
condition for fast thermalization assumed in the standard rate calculation and show that it is generally violated 
in commonly studied field theories. The ensuing 
non-linear non-equilibrium dynamics tends to suppress 
the decay rate below the prediction of the thermal theory. 

We argue, however, that at much lower temperatures, 
$E_b/T\gtrsim 100$, the non-thermal effects become irrelevant and the rate 
(\ref{GammaAffleck}) gets recovered, up to small perturbative corrections. 
Our argument applies to non-dissipative systems. It 
uses the Liouville theorem combined with the time-reversal invariance and connects the rate of false vacuum decay to the probability that a collapsing critical bubble will dynamically ``turn around" and start expanding. Extension of the argument to systems with dissipation and further investigation of the critical bubble dynamics
will be reported elsewhere \cite{upcoming}.

\section{Setup}
\label{sec:setup}

We consider a real scalar field in 
$(1+1)$ dimensions with the action
\be \label{S}
    S=\int\diff t \, \diff x\left( \frac{(\d_\mu\phi)^2}{2} - \frac{m^2\phi^2}{2} +\frac{\l \phi^4}{4} \right) \, ,
\ee
where $\l>0$. The false vacuum is located at $\phi=0$, and the true vacuum corresponds to the run-away ${\phi\to\pm\infty}$. The choice of the quartic potential is convenient since it allows us to determine all quantities entering the Euclidean prediction for the rate analytically. 
However, our conclusions do not rely on this choice.

In the theory (\ref{S}), the critical bubble profile, its energy and the growth rate of its unstable mode are:
\be \label{bubble}
    \phi_b(x) = \sqrt\frac{2}{\l} \cdot \frac{m}{\ch{mx}}\, , \quad
    E_b = \frac{4m^3}{3\l} \, , \quad \omega_- = \sqrt{3} \, m\, .
\ee
Evaluating the critical bubble contribution to the free energy (see Appendix~\ref{app:rate}) and substituting it into the Euclidean formula (\ref{GammaAffleck}), one obtains the nucleation rate:
\be \label{G_E2}
    \G_E= \frac{6m^2}{\pi}\sqrt{\frac{E_b}{2\pi T}}\, \e^{-E_b/T}\, .
\ee
Below, we compare this expression with the results of real-time numerical simulations.

We discretize the action (\ref{S}) on a periodic spatial lattice with step $a$ and length $L$ using the second-order finite-difference approximation for spatial derivatives. This leads to multi-dimensional Hamiltonian dynamics, whose classical equations are evolved using the $4^{\rm th}$-order operator-splitting pseudo-spectral method \cite{Pirvu:2024nbe}. Most runs are performed with $a\simeq0.012/m$, $L=100/m$, and time step $\Delta t\simeq 0.8a$. We have verified the convergence of our results by varying the lattice parameters in the ranges $ma\in [5\cdot 10^{-3},4\cdot 10^{-2}]$, $mL\in [50,400]$, $\Delta t/a\in [0.4,0.8]$. The simulations were also cross-checked with an independent code \cite{Pirvu:2023plk} based on a $10^{\rm th}$-order Gauss-Legendre pseudo-spectral scheme, and choosing $ma=0.04$, $mL\in [80,100]$ and $\Delta t\simeq 0.17a$.

The initial conditions for the simulations correspond to the thermal equilibrium around the false vacuum $\phi=0$ with the temperature $T$. We prepare the initial state using the Hamiltonian Monte Carlo (HMC) method \cite{Neal:2011mrf}. This differs from our previous work \cite{Pirvu:2024nbe} where the initial state was taken to be Gaussian, with the leading effect of the field interactions taken into account by renormalizing the mass of the field. One could expect the difference between the Gaussian and exact thermal state to be small at low temperature, $\l T/m^3\ll 1$. We find, however, that it has a significant (about a factor 1.5) effect on the false vacuum decay rate. Details of our HMC implementation and the results for Gaussian initial state 
can be found in 
Appendix~\ref{app:hmc}.

\section{Survival probability under Hamiltonian evolution}
\label{sec:Psurv}

\begin{figure}[t]
	\center{	
		\begin{minipage}[h]{0.99\linewidth}
			\center{\includegraphics[width=1.0\linewidth]{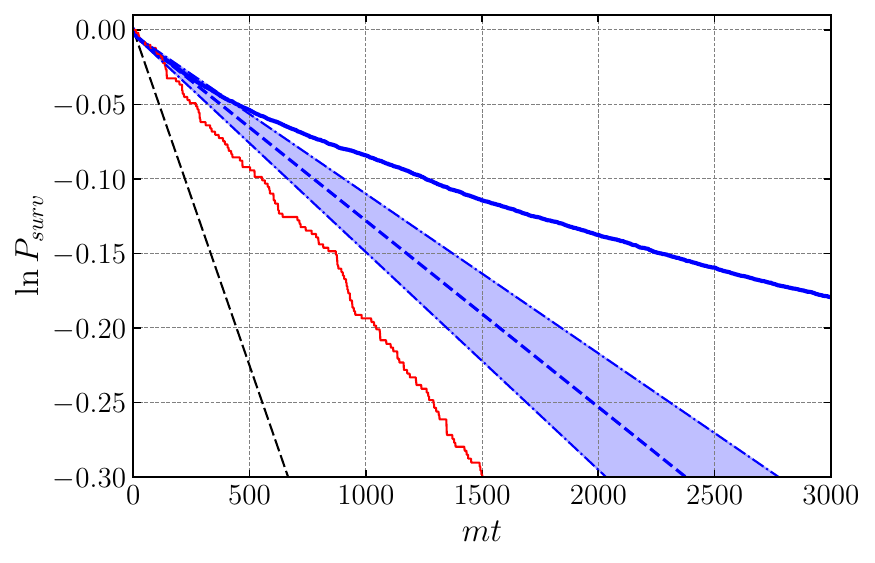}}
		\end{minipage}
	}
  \caption{{\it Blue thick solid:} Survival probability in real-time simulations of Hamiltonian dynamics for $\lambda T/m^3=0.1$.
  {\it Blue thick dashed:} Straight line with the slope equal to the best fit value of the derivative of the previous curve at $t=0$. 
  {\it Blue band:} Statistical uncertainty of the fit. 
   {\it Red thin solid:} Survival probability from Langevin dynamics (Eq.~(\ref{LangEq})) at the same temperature and with $ \eta = 0.01m $. Wiggles in the curve correspond to Poisson fluctuations.
   {\it Black thin dashed:} Prediction of the Euclidean theory.
  }\label{fig:Psurv}
\end{figure}

We generate an ensemble of simulations with temperature $T$ and monitor them until they decay into the true vacuum. At each moment of simulation time $t$, we count the number of configurations that have not yet decayed. The {\it survival probability} $P_{surv}(t)$ is then defined as the ratio of this number to the total initial number of configurations in the ensemble. This measurement is repeated for several choices of temperature in the range $0.08 \leqslant \l T/m^3 \leqslant 0.11$. A typical result is shown by the upper curve in Fig.~\ref{fig:Psurv}.

For decays obeying the exponential distribution, the survival probability follows the law
\be \label{Psurv}
    \ln P_{surv}(t)=\const-\G L\cdot t \;.
\ee
The thin black dashed line in Fig.~\ref{fig:Psurv} shows such a curve, using the Euclidean prediction (\ref{G_E2}) for the rate. We see a clear discrepancy between the prediction and the real-time data, which calls for an explanation.

{\bf\emph{Flattening of ${\boldsymbol \ln}\, \boldsymbol{P_{surv}}$:`Classical Zeno effect'}} --- The measured survival curve in Fig.~\ref{fig:Psurv} is not straight: it flattens out as time increases, implying a decrease of the decay rate. The reason for this behavior lies in the dynamics of bubble nucleation. The critical bubble is composed of long Fourier modes with wavenumbers ${k\lesssim m}$, while most of the field energy is stored in shorter modes. The latter provide a thermal bath for the former. However, the energy exchange between different modes is inefficient \cite{Boyanovsky:2003tc}. In the model at hand, it is dominated by $2\leftrightarrow 4$ and $3\leftrightarrow 3$ scattering.\footnote{$2\leftrightarrow 2$ scattering preserves the energy distribution due to $(1+1)$-dimensional kinematics.} The corresponding thermalization time is estimated as (see Appendix~\ref{app:therm}),\footnote{Its parametric form can be found on dimensional grounds by first restoring $\hbar$ and then requiring that it drops off $t_{th}$ in the classical limit.}
\be \label{t_therm}
    t_{th}\sim\frac{(2\pi)^3}{m}\left(\frac{m^3}{\l T}\right)^4 \; .
\ee
For the temperatures in our simulations $t_{th}\gtrsim 10^6/m$ which is longer than the typical decay time $t_{dec} \sim (\Gamma L)^{-1}\sim 10^4/m $. The initial power contained in the long modes is then essentially preserved for each individual simulation and controls its lifetime. A simulation which, due to a statistical fluctuation, has a higher long-mode power decays faster, while the one with lower power lives longer.

\begin{figure}[t]
	\center{	
		\begin{minipage}[h]{0.95\linewidth}
			\center{\includegraphics[width=1.0\linewidth]{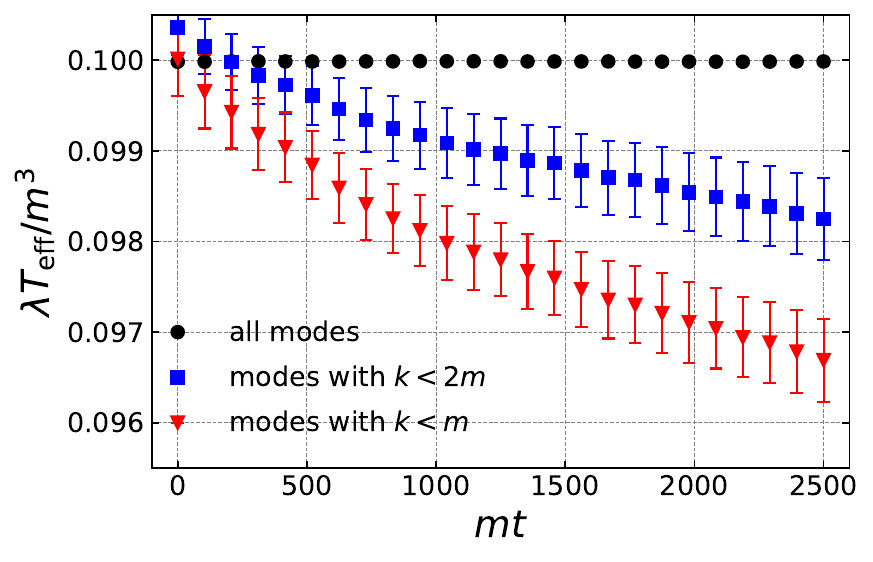}}
		\end{minipage}
	}
  \caption{Effective temperature of long modes Eq.~(\ref{eq:Teff}) for ${k_*=m}$ and $k_*=2m$, averaged over the surviving configurations at time $t$. The temperature computed using all modes is also shown and is time-independent.
  }
	\label{fig:Zeno}
\end{figure}

This, in turn, biases the statistical properties of the surviving ensemble. As the time goes on, the average long-mode power decreases. The effect is apparent in Fig.~\ref{fig:Zeno}, where we plot the effective temperature of long modes,
\be
\label{eq:Teff}
T_{\rm eff}=\langle |\tilde\pi_j|^2\rangle_{k_j<k_*}\;.
\ee
Here $\tilde \pi_j$ is the Fourier component of the canonical momentum $\pi=\dot\phi$ with wavenumber $k_j$ (see Appendix~\ref{app:hmc} for the precise definition); the average is taken over the modes with wavenumbers below $k_*$ and over the surviving configurations at time $t$. The effective temperature decreases by a few per cent during the run, enough to considerably suppress the bubble nucleation rate. Rather unexpectedly, the decay happens to be non-Markovian: a system that is observed not to decay within a given time has a lower chance of decaying in the future. This is reminiscent of the Zeno effect, which allows one to freeze the evolution of a quantum system by measurements. We stress, however, that in our case, the effect is purely classical.

The drift in the rate is expected to disappear when the decay is slow enough so that the condition ${t_{dec}\gg t_{th}}$ is satisfied. This condition is likely fulfilled in most cosmological settings. On the other hand, whether it holds in laboratory experiments, especially in those using $(1+1)$-dimensional systems, is less evident. In this case, the classical Zeno effect should be taken into account.

{\bf\emph{Unbiased rate}} --- For quantitative comparison with the Euclidean prediction (\ref{G_E2}), we measure the slope $\diff\ln P_{surv}/\diff t$ at $t\to 0$ corresponding to the rate in the initial unbiased ensemble. We use an extrapolation procedure to increase the accuracy. The probability curve is split into small, approximately linear segments, and the slope of each segment is measured. 
The logarithms of the slopes thus obtained are fitted with a quadratic function of time whose value at $t=0$ yields the logarithm of the unbiased rate $\Gamma$.
The details of this procedure are described in Appendix~\ref{app:extrap}.
The error introduced by the procedure is dominated by statistical uncertainty. Repeating it at different temperatures, we obtain the unbiased rate $\G(T)$, which we compare with the theoretical prediction~(\ref{G_E2}). 

\begin{figure}[t]
\center{
		\begin{minipage}[h]{0.99\linewidth}
			\center{\includegraphics[width=0.99\linewidth]{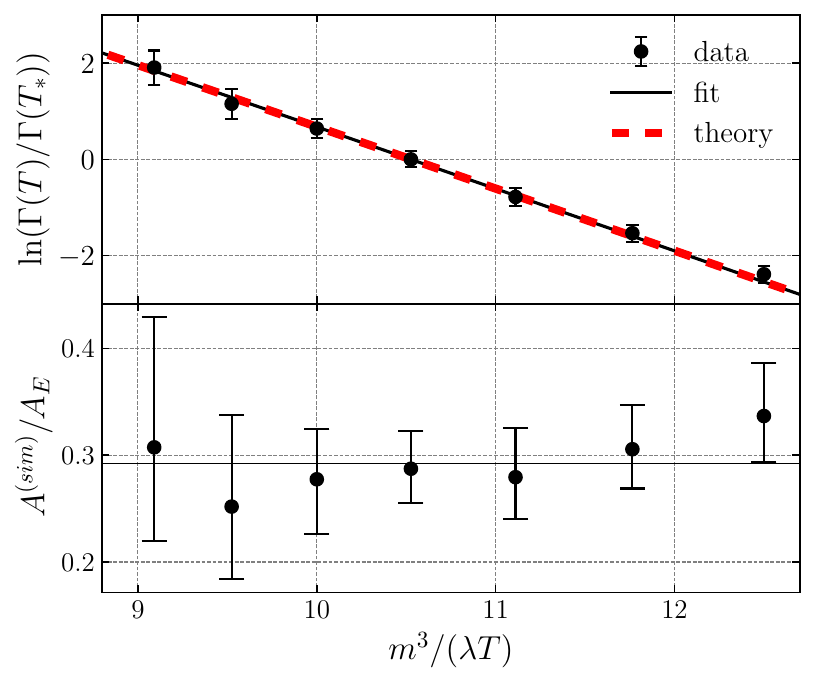}}
		\end{minipage}
	}
  \caption{\textbf{Top:} Measurement of the exponential suppression of the decay rate. The unbiased rate normalized at a pivot point $\lambda T_*/m^3=0.095$ is shown as function of the inverse temperature (black dots). The black solid line corresponds to Eq.~(\ref{log_G(T)}) with the best fit value of $B$. The red dashed line is the prediction of Euclidean theory. \textbf{Bottom:} 
  Prefactor extracted from real-time simulations $A^{(sim)}$ vs. Euclidean prediction $A_E$. The horizontal line shows the average.}
\label{fig:rate}
\end{figure}

We fit $\Gamma(T)$ 
by the expression,
\be \label{log_G(T)}
    \ln\G(T) =- \frac{1}{2}\ln T + \ln A - \frac{B}{T} \;,
\ee
with free parameters $A$ and $B$. The first term on the right accounts for the temperature dependence of the prefactor predicted by Eq.~(\ref{G_E2}).
We first measure the exponential suppression part of the rate.
We eliminate the constant $A$ by taking the ratio $\G(T)/\G(T_*)$ with $T_*$ from the middle of the interval and determine the slope $B$. The fit is shown in the top panel of Fig.~\ref{fig:rate}. It gives $B/E_b=0.96\pm 0.04$ with $E_b$ from Eq.~(\ref{G_E2}), consistent with the predicted Boltzmann suppression. Note that the bubble energy is computed using the bare field mass, without any  
thermal corrections, cf.~\cite{Alford:1993zf,Gleiser:1993hf, Alford:1993ph}.

Next, we measure the prefactor. We fix $B=E_b$ in Eq.~(\ref{log_G(T)}) and extract $A$ at different values of temperature. The ratio of the result to the Euclidean prediction of Eq.~(\ref{G_E2}) is shown in the bottom panel of Fig.~\ref{fig:rate}. The measured prefactor $A^{(sim)}$ is smaller than the prediction $A_E$: the fit gives $A^{(sim)}/A_E=0.29\pm0.02$.\footnote{A two-parameter fit of $\ln\Gamma(T)$ by the expression in Eq.~(\ref{log_G(T)}) yields a consistent result.}\;\footnote{This value is a factor $\sim 2$ higher than reported in \cite{Pirvu:2024nbe}. The difference is due to two reasons. First, in the present paper we prepare the initial state using HMC, instead of the Gaussian approximation. This increases the measured rate by a factor $\sim 1.5$, see Appendix~\ref{app:hmc}. Second, we use a more conservative extrapolation procedure for the determination of the unbiased rate, see Appendix~\ref{app:extrap}.}
This discrepancy cannot be attributed to two-loop corrections, which are expected to affect the prefactor only at the $\lambda T/m^3\sim 10\%$ level.
The prefactor is temperature-independent within the error bars, though a slight upward trend of the central values can be noticed in the right part of the plot corresponding to the lowest temperatures.

\section{Decay with an external heat bath}
\label{sec:Lang}

To investigate the system further, we artificially reduce its thermalization time by coupling it to an external heat bath. This is implemented by promoting the equation of motion to the Langevin equation,
\be \label{LangEq}
    \ddot\phi + \eta\dot\phi - \phi^{\prime\prime} + m^2\phi - \l\phi^3 = \xi \; ,
\ee
where $\eta$ is the friction coefficient and $\xi(t,x)$ is the white noise, whose amplitude is fixed by the fluctuation-dissipation theorem:
\be \label{Noise}
    \langle\xi(t,x)\xi(t^{\prime},x^{\prime})\rangle = 2\eta T\,\delta(t-t^{\prime})\delta(x-x^{\prime}) \, .
\ee
We solve this equation numerically using a $3^{\rm rd}$-order stochastic pseudo-spectral operator-splitting scheme \cite{Pirvu:2024nbe,Shkerin:2025hui}. 
Note that the evolution becomes insensitive to the initial conditions after the thermalization time $t_{th}\sim \eta^{-1}$. Thus, as soon as $\eta\gg \Gamma L$, we take the initial state prepared according to the Gaussian distributions, which is computationally cheaper (see Appendix~\ref{app:hmc} for details). Otherwise, if $\eta\lesssim \Gamma L$, we use the HMC method, as for the Hamiltonian evolution. 

\begin{figure}[t]
\center{
		\begin{minipage}[h]{0.99\linewidth}
			\center{\includegraphics[width=0.99\linewidth]{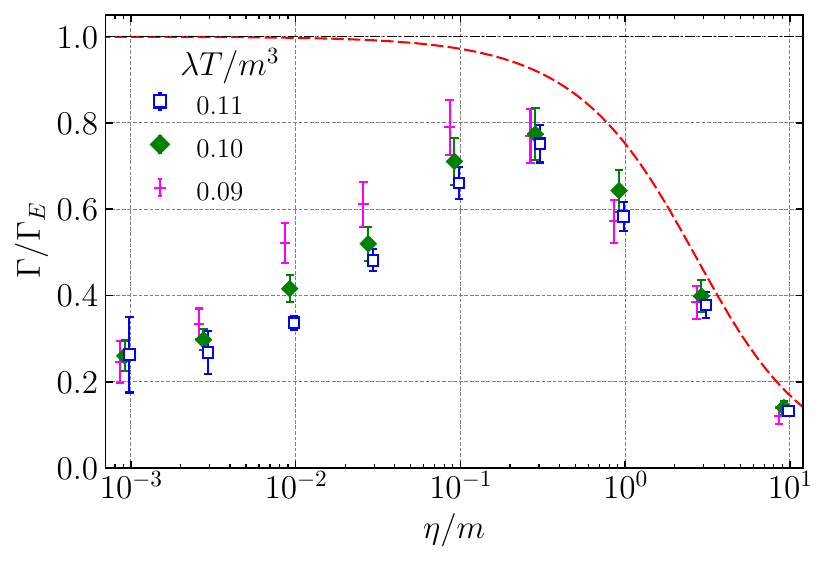}}
		\end{minipage}
	}
  \caption{Decay rate measured in simulations as function of the dissipation coefficient $\eta$ for several values of temperature. The rate is normalized by the prediction of Euclidean theory, Eq.~(\ref{G_E2}). The red dashed line shows the prediction of classical statistical theory, Eq.~(\ref{GammaLanger}).
  }
	\label{fig:pref_Eta}
\end{figure}

We observe that if $\eta\gtrsim 10^{-3}m$ the survival probability curves $\ln P_{surv}(t)$ follow straight lines, modulo random fluctuations (see the red curve in Fig.~\ref{fig:Psurv} corresponding to $\eta=0.01m$).
This is expected, since now $t_{th}<t_{dec}$ and the classical Zeno effect is absent. However, their average slope is still less than the Euclidean prediction.
The measured rate $\Gamma(\eta, T)$ grows as we increase $\eta$ until it reaches a maximum at $\eta\sim 0.3m$ where it deviates from $\Gamma_E(T)$ by only $20\%$. At larger $\eta$, it decreases again, consistent with Eq.~(\ref{GammaLanger}). This behavior is shown in Fig.~\ref{fig:pref_Eta}.

\section{Violation of thermal equilibrium}
We see from Fig.~\ref{fig:pref_Eta} that the decay rate is closest to Eq.~(\ref{GammaLanger}) at $\eta\gtrsim 0.1m$ when the thermalization time is comparable to the inverse mass, and deviates from it at smaller $\eta$ when the thermalization time is longer. This suggests associating the deviation with the lack of thermal equilibrium during the bubble nucleation process, whose dynamical time scale is set by $\omega_-=\sqrt{3} m$ (see Eq.~(\ref{bubble})). In simple mechanical systems with one  degree of freedom 
the applicability of Eq.~(\ref{GammaLanger}) is known to require~\cite{Hanggi:1990zz} 
\be \label{etaLanger}
    {\eta/\omega_-}>{T/E_b}\;,
\ee
where $E_b$ is the height of the barrier. Otherwise, Eq.~(\ref{GammaLanger}) overestimates the rate. The condition (\ref{etaLanger}) ensures 
that the states in the vicinity of the barrier are populated according to the 
Boltzmann distribution. The standard derivation of Eq.~(\ref{GammaLanger}) for multidimensional systems \cite{Langer:1969bc} also assumes a thermal distribution of states in the phase space close to the critical bubble and breaks down when the condition (\ref{etaLanger}) is violated \cite{Pirvu:2024nbe} (see also \cite{Ekstedt:2022tqk}).  
Note that for fixed $\eta$, this condition is always satisfied at a low enough temperature. Thus, the rate $\Gamma(\eta, T)$ is expected to approach Eq.~(\ref{GammaLanger}) from below at $T\to 0$ for any ${\eta>0}$. This is consistent with the trend exhibited by the simulation data in the range $10^{-3}m\lesssim \eta\lesssim 0.1m$ (see Fig.~\ref{fig:pref_Eta}), though the measured rate is still far from the limit for the explored temperatures.

Since the condition (\ref{etaLanger}) ensures efficient thermalization, it is sufficient for the validity of the equilibrium false vacuum decay theory. However, it is 
formulated for systems coupled to external heat bath and cannot be directly applied for a Hamiltonian system with $\eta=0$. For such systems, one can heuristically replace $\eta$ in Eq.~(\ref{etaLanger}) with the inverse thermalization time, so that the condition for thermal equilibrium becomes,  
\be \label{criterion}
    t_{th}<\frac{E_b}{T\omega_-}\;.
\ee
Comparison with Eq.~(\ref{t_therm}) shows that 
this condition is violated 
in the Hamiltonian evolution of the theory (\ref{S}), signaling a breakdown of thermal equilibrium during the critical bubble nucleation.

This violation is not tied to the peculiarities of the $(1+1)$-dimensional model. 
For example, in classical $\lambda\phi^4$ theory in $(3+1)$ dimensions the thermalization time scales as $t_{th} \propto m/(\lambda T)^2$ (see Appendix~\ref{app:therm}), whereas $E_b\propto m/\lambda$ and thus Eq.~(\ref{criterion}) is always violated as long as $T<E_b$, i.e. as long as the vacuum decay is exponentially suppressed.
The same is typically true for any weakly coupled theories, unless they feature a strong hierarchy of couplings, or contain a large number of fields. We conclude that nucleation of critical bubbles is an essentially non-equilibrium process.

\section{Recovery of equilibrium rate}
\label{sec:Cond}

We now argue that despite the violation of thermal equilibrium discussed above, the false vacuum decay rate in a Hamiltonian theory is still given by Eq.~(\ref{GammaAffleck}) at low enough temperature. To this aim, we present a derivation of the false vacuum decay rate which does not rely on thermal population of states in the vicinity of the critical bubble.

Consider the evolution of the field from an initial configuration $\{\phi_1(x),\pi_1(x)\}$ describing a thermal excitation around the false vacuum at time $t_1=0$ to a state containing the critical bubble at time $t_2$. The time $t_2$ is assumed to be longer than the typical dynamical time of the system, $t_2> 2\pi/m$, but shorter than the false vacuum decay time, so that the depletion of the ensemble due to the classical Zeno effect can be neglected.
The final state can be represented as the sum of the critical bubble profile $\phi_b(x)$ and perturbations on top of it. These contain an unstable mode with the mode function $\vf_{0,0}(x)$ (see Appendix~\ref{app:rate}) and stable perturbations\footnote{We include here the zero mode corresponding to the bubble translations.}  which we collectively denote by $\phi_+(x)$. The field and its canonical momentum at time $t_2$ are thus written as, 
\bseq
\label{finstate}
\begin{align}
& \phi_2(x)=\phi_b(x)+\chi_-\vf_{0,0}(x)+\phi_+(x) \;, \\
& \pi_2(x)=p_-\vf_{0,0}(x)+\pi_+(x) \;,
\end{align}
\eseq
where $\chi_-$, $p_-$ are the amplitude of the unstable mode and its momentum. Recall that the critical bubble represents the saddle-point configuration of the energy, dividing the phase space into basins of attraction of the false and true vacua. Without loss of generality, we can choose $\chi_-<0$ ($\chi_->0$) to lie on the false (true) vacuum side.

The fraction of configurations in the initial ensemble that cross the divide from $\chi_-<0$ to $\chi_->0$ within the time interval from $t_2$ to $t_2+dt$ gives the probability of the false vacuum decay in this time interval, 
\be 
\label{Pdec}
\begin{split}
P_{\rm dec}[t_2,t_2+dt]=\frac{1}{Z}&\int [d\phi_1(x)d\pi_1(x)]\, \e^{-H/T}\\
&\times \Theta_{\rm cross}[t_2,t_2+dt;\phi_1,\pi_1]\;,
\end{split}
\ee
where
\be
Z=\int [d\phi_1(x)d\pi_1(x)]\, \e^{-H/T}
\ee
is the false vacuum partition function, $H$ is the system's Hamiltonian, and $\Theta_{\rm cross}$ stands for the condition of crossing the critical bubble by the final configuration. Note that due to the deterministic nature of the evolution, $\Theta_{\rm cross}$ is a well-defined function of the initial data.

The deterministic map between the initial and final states allows us to transform the integral in (\ref{Pdec}) into the integral over the final data,
\be 
\label{Pdec1}
\begin{split}
P_{\rm dec}&[t_2,t_2+dt]=\frac{1}{Z}\int [d\chi_-dp_-d\phi_+(x)d\pi_+(x)]\, \e^{-H/T}\\
&\times J\, \theta(-\chi_-)\theta(\chi_-+p_-dt)\,
\Theta_{\rm false}[\chi_-,p_-,\phi_+,\pi_+]\;,
\end{split}
\ee
where 
\be
J=\bigg|\frac{\d (\phi_1,\pi_1)}{\d(\chi_-,p_-,\phi_+,\pi_+)}\bigg|
\ee
is the Jacobian of the transformation and $\Theta_{\rm cross}$ takes now an explicit form with the usual Heaviside $\theta$-fun\-ctions. However, we have to add a new condition, represented by $\Theta_{\rm false}$, that the field trajectory starts from the false vacuum. 

A key observation is that by Liouville theorem $J=1$. The integral over $\chi_-$ can then be taken explicitly and we arrive at the decay rate,
\be
\label{GammaTrue}
\begin{split}
\Gamma=&\frac{P_{\rm dec}(t_2,t_2+dt)}{L dt}\\
=&\frac{1}{ZL} \int [dp_-d\phi_+(x)d\pi_+(x)]\, p_-\theta(p_-)\,\e^{-H/T}\\
&\qquad\times \Theta_{\rm false}[\chi_-=0,p_-,\phi_+,\pi_+]\;.
\end{split}
\ee
If $\Theta_{\rm false}$ were equal to unity, we would exactly recover the equilibrium thermal flux through the saddle point leading to the rate (\ref{GammaAffleck}) \cite{Affleck:1980ac}. But in general $\Theta_{\rm false}$ can differ from $1$ if there are configurations ${p_-,\phi_+(x),\pi_+(x)}$ that do not correspond to any initial data starting in the false vacuum. 
Under the reversed time evolution they correspond to field trajectories that start as a perturbation on the critical bubble, evolve towards the false vacuum, but instead of dissolving into it, ``turn around" and cross the critical bubble again in the direction of the true vacuum.

At first sight, existence of such turn-around trajectories may appear surprising. However, we have explicitly confirmed it in simulations. A detailed analysis will be presented in \cite{upcoming}, here we briefly summarize the results. The turn-around trajectories split into two classes: ``prompt" and ``delayed". 
The prompt trajectories do not leave the vicinity of the critical bubble before going to the ``wrong" side. They result from the interactions between the unstable mode and the positive modes which reverse the sign of $p_-$ before the configuration has time to escape from the saddle point. 
These trajectories are expected to give perturbative (two-loop) corrections to the decay rate. On the other hand, the delayed configurations evolve all the way towards the false vacuum, start oscillating around it, but then rebound and escape to the true vacuum.\footnote{In the model (\ref{S}) the trajectories describing collapse of a critical bubble with $\phi(x)>0$ typically go over to the opposite vacuum at $\phi\to-\infty$ after half of an oscillation. However, a non-negligible fraction of trajectories completes one or several full oscillations and go to $\phi\to+\infty$. We have also checked the existence of delayed turn-around trajectories in field theories with asymmetric potential and single true vacuum.} 
They arise from a constructive interference between the large coherent field oscillations resulting from the critical bubble collapse and random thermal fluctuations. Accounting for them provides non-perturbative corrections to the rate.  

Returning to Eq.~(\ref{GammaTrue}), we can write the decay rate as 
\be 
\label{RateRel}
\Gamma=\Gamma_E\,( 1-R )\;,
\ee
where
\be
R=\frac{\int [dp_-d\phi_+d\pi_+]\,|p_-|\,\theta(-p_-)\,\e^{-H/T}\,(1-\Theta_{\rm false})}
{\int [dp_-d\phi_+d\pi_+]\,|p_-|\,\theta(-p_-)\,\e^{-H/T}}
\ee
is the probability for a trajectory describing the collapse of the critical bubble to turn around and escape to the true vacuum. Note that in writing this expression we have used the time reversal invariance and flipped the sign of $p_-$. In the limit of very low temperature, $T\to 0$, we expect the probability $R$ to vanish: in this limit we simply have a collapse of the critical bubble, without any thermal fluctuations that could reverse it. Then $\Gamma\approx \Gamma_E$ implying that the rate is given by the equilibrium formula. On the other hand, our results from pervious sections indicate that $R\sim 0.7$ for $E_b/T\sim 10$.

\begin{figure}[t]
\center{
		\begin{minipage}[h]{0.99\linewidth}
			\center{\includegraphics[width=0.99\linewidth]{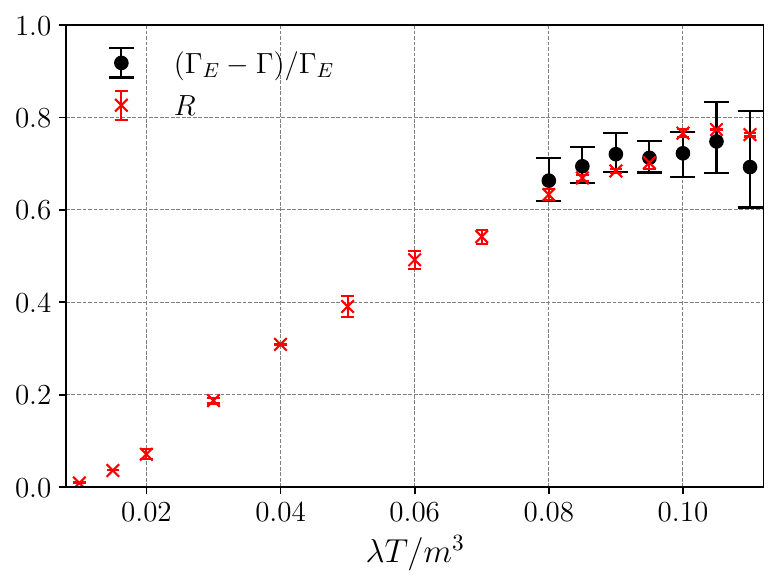}}
		\end{minipage}
	}
  \caption{Turn-around probability $R$ as function of temperature ({\it red crosses}) and the relative difference $(\Gamma_E-\Gamma)/\Gamma_E$ between the measured and equilibrium false vacuum decay rates ({\it black dots}). }
	\label{fig:returns}
\end{figure}

It is straightforward to measure $R$ directly in simulations by initiating a suite of configurations with thermal fluctuations superimposed on a collapsing critical bubble. To this aim, we decompose the perturbations of the critical bubble into eigenmodes and populate the positive modes using the Gaussian distribution for their amplitudes and momenta, see Appendix~\ref{app:hmc}.\footnote{Since $R$ is not exponentially suppressed, the error introduced by using the Gaussian approximation to the exact thermal distribution is negligible.} 
The amplitude of the negative mode $\chi_-$ is fixed to zero, while its momentum is taken to be negative and sampled with the weight $|p_-|\e^{-p_-^2/2T}$. We then evolve these configurations for a time $t$ equal to a few dynamical times, $t\simeq 200/m$, and count the fraction of configurations that decay to the true vacuum. Note that the evolution time is chosen to be much shorter than $(\Gamma L)^{-1}$, so that the contribution of vacuum decays unrelated to the critical bubble collapse is negligible. We run $10^3$ simulations for each value of the temperature. The results are shown in Fig.~\ref{fig:returns}. We see that at
$0.08\leqslant \lambda T/m^3\leqslant 0.11$ they agree well with the relative difference $(\Gamma_E-\Gamma)/\Gamma_E$ between the decay rate measured in Sec.~\ref{sec:Psurv} and the equilibrium value. At $\lambda T/m^3<0.08$, where a direct measurement of $\Gamma$ is not available, the turn-around probability $R$ decreases and becomes less than $0.01$ at $\lambda T/m^3\lesssim 0.01$.

The above discussion assumes that a field trajectory crossing the critical bubble from $\chi_-<0$ to $\chi_->0$ necessarily ends up in the true vacuum. We have verified that this is indeed the case in the model (\ref{S}). However, this need not be true in general and field trajectories can cross the divide multiple times, cf.~\cite{Moore:2000jw}. Modifications to Eq.~(\ref{RateRel}) that include this possibility, as well as the generalization to stochastic systems will be considered elsewhere \cite{upcoming}.

\section{Discussion}
\label{sec:disc}

Using direct numerical simulations, we have found suppression of thermal false vacuum decay rate compared to the classic results (\ref{GammaAffleck}), (\ref{GammaLanger}). This suppression arises at the level of prefactor and only at moderately low temperatures, $T\sim 0.1 E_b$. In the limit $T\to 0$ the discrepancy disappears. Nevertheless, it is conceptually important since it reveals non-equilibrium dynamics of critical bubble nucleation. 
Thermal false vacuum decay turns out not to differ fundamentally from the decay under general non-equilibrium conditions \cite{Kofman:1995fi, Khlebnikov:1998sz, Gleiser:2004iy, Gleiser:2007ts, Pirvu:2023plk, Hiscock:1987hn, Berezin:1990qs, Gregory:2013hja, Tetradis:2016vqb, Gorbunov:2017fhq, Mukaida:2017bgd, Kohri:2017ybt, Hayashi:2020ocn, Shkerin:2021zbf, Shkerin:2021rhy, Strumia:2022jil, Hu:2026udq}, even if in the latter cases the rate is altered already at the level of the exponential suppression.

We have presented a derivation of the false vacuum decay rate that does not rely on equilibrium conditions in the part of the phase space close to the critical bubble. This has allowed us to connect the suppression of the prefactor to the probability that a collapsing critical bubble
turns around into an expanding one due to thermal fluctuations.
Our derivation assumes deterministic evolution and absence of field trajectories with multiple crossings of the divide between the false and true vacua. Generalization to systems with multiple crossings and stochastic dynamics is left for future \cite{upcoming}.

Our findings are consistent with the results of Ref.~\cite{Hirvonen:2025hqn} which studied the time-dependent decay rate of an ensemble of systems (\ref{S}) initially distributed with Boltzmann weights throughout the whole basin of attraction of the false vacuum, including the states close to the critical bubble. That work found that the rate starts from the initial value (\ref{G_E2}), but then quickly drops by a factor $\sim 2$ within the dynamical time due to the depletion of states in the vicinity of the critical bubble. Our results indicate that at times $t\sim 100/m$ the ensemble will reach a steady state with the rate given by Eq.~(\ref{RateRel}). The simulations in \cite{Hirvonen:2025hqn} are performed on shorter time scales, precluding a detailed comparison. We plan to return to such comparison in future.

On even longer time scales the decay rate further decreases due to the statistical biasing of the simulated ensemble by the depletion of 
configurations with higher power in long Fourier modes. 
This depletion is not compensated by the energy exchange between the long and short modes, which is too slow in the system (\ref{S}).
We called this long-time decrease of the decay rate classical Zeno effect. 
The energy exchange is typically more efficient in dimensions higher than $(1+1)$, so it is unclear if 
the classical Zeno effect is important for high energy phenomenology. It would be interesting to study its relevance for cold atom experiments.

In the present paper we focused on the purely classical false vacuum decay relevant at high temperatures. It would be interesting to extend the study by including the quantum corrections. One may further wonder if non-equilibrium dynamics can play any role in the deeply quantum regime of very low or zero temperature when the false vacuum decay proceeds by tunneling. A promising toolset for studying these questions is provided by the semiclassical methods developed in 
 \cite{miller1974classical, Rubakov:1992ec, Kuznetsov:1997az, Bonini:1999kj, Bezrukov:2003er, Bezrukov:2003tg, Levkov:2004ij, Levkov:2007yn, Levkov:2008csa, Demidov:2011dk, Demidov:2015bua, Demidov:2015nea, Shkerin:2021zbf}.

Finally, our results also motivate revisiting the real-time dynamics of other non-perturbative processes, such as e.g. sphaleron transitions in the early universe \cite{Kuzmin:1985mm, Rubakov:1996vz}, suggesting that it can hide non-equilibrium features.

{\bf\emph{Acknowledgments}} --- We are grateful to Oliver Gould, Joonas Hirvonen and Marina Marinkovic for fruitful discussions. 
We thank Mustafa Amin, Asimina Arvanitaki, Claudia Cornella, Marco Costa, Ruth Gregory, Junwu Huang, Matthew Johnson, Alexander Kayssi, Juraj Klaric, Andrew Kovachik, Sung-Sik Lee, Alexander Penin, Maxim Pospelov, Jury Radkovski, Kam To Billy Sievers, Mikhail Shaposhnikov and Andrei Zelnikov for stimulating interest. Research at Perimeter Institute is supported in part by the Government of Canada through the Department of Innovation, Science and Economic Development Canada and by the Province of Ontario through the Ministry of Colleges and Universities. The work of SS is supported by the Natural Sciences and Engineering Research Council (NSERC) of Canada.

\appendix

\section{Euclidean calculation of the decay rate}
\label{app:rate}

\noindent Here we compute the imaginary part of the false vacuum free energy $F$ 
entering Eqs.~(\ref{GammaAffleck}), (\ref{GammaLanger}) in the model~(\ref{S}). We start with the partition function,
\be \label{Zgen}
    \begin{split}
        Z =\int [\diff\phi] \exp\bigg\{-\int \diff\tau \diff x\,
        \bigg(\frac{(\d_\mu\phi)^2}{2}+\frac{m^2\phi^2}{2}
        -\frac{\l\phi^4}{4}\bigg)\bigg\},
    \end{split}
\ee
where the path integral runs over Euclidean configurations with period $1/T$. Besides the vacuum $\phi=0$, the integral has two non-trivial saddle points: the critical bubble $\phi_b(x)$ and its reflection $-\phi_b(x)$. They correspond to decays towards $\phi=\pm\infty$ and together contribute
\be \label{Zsph}
    Z_b=2\cdot\bigg(\frac{{\cal D}}{{\cal D}^{(0)}}\bigg)^{-1/2}\e^{-S[\phi_b]}\;.
\ee
Here $S[\phi_b]=E_{b}/T$ is the bubble action, and ${\cal D}$ is the determinant of small fluctuations around it, normalized by the vacuum determinant ${\cal D}^{(0)}$. 
It is well-known that ${\cal D}$ is negative due to the presence of the unstable mode,
hence the right-hand side of Eq.~\eqref{Zsph} is imaginary. This gives an imaginary contribution to the free energy,
\be \label{ImF}
    \Im F \equiv -T\,\Im \ln Z
    = T\cdot 2 \cdot \frac{1}{2} \bigg| \frac{\D}{\D^{(0)}} \bigg|^{-1/2} \e^{-E_b/T}\;,  
\ee
where we have included the factor $1/2$ coming from the integration over the negative mode \cite{Coleman:1978ae}.

The determinant ${\cal D}$ is the product over eigenvalues $\a_I$ of the linearized equation for perturbations:
\be \label{eigen1}
    -\Box\vf_I+ \big(m^2-3\l\phi_b(x)\big)\vf_I=\a_I\,\vf_I\;.
\ee
The functions $\vf_I$ are periodic in the Euclidean time. Since the bubble is time-independent, we can separate the variables:
\be \label{separ}
    \vf_I(\tau,x)=\e^{-i\omega_n \tau}\vf_{n,k}(x)\;,
\ee
where $\o_n=2\pi n T$ is the $n^{\rm th}$ Matsubara frequency and the spatial functions satisfy
\be \label{eigen2}
    {\cal O}(\mu_n)\,\vf_{n,k}= \alpha_{n,k}\,\vf_{n,k}\;, \qquad \mu_n^2=\omega_n^2+m^2\;,
\ee
with the operator
\be \label{Omu}
    {\cal O}(\mu)=-\d^2_x+\mu^2-\frac{6m^2}{\ch^2{mx}}\;.
\ee
The negative mode lies in the $n=0$ sector and reads:
\be \label{NegMode} 
    \vf_{0,0}(x)\propto \frac{1}{\ch^2 mx} \;, \qquad \a_{0,0}\equiv-\omega_{-}^2 = -3m^2 \;.
\ee
The determinant ratio factorizes into a product over the Matsubara sectors:
\be \label{factor}
    \frac{\D}{\D^{(0)}}=\prod_{n=-\infty}^\infty \frac{\D_n}{\D^{(0)}_n} = \frac{\D_0}{\D^{(0)}_0}\prod_{n=1}^\infty \left(\frac{\D_n}{\D^{(0)}_n}\right)^2\;,
\ee
where $\D_n$ and $\D_n^{(0)}$ are determinants of 
${\cal O}(\mu_n)$ and ${\cal O}^{(0)}(\mu_n)=-\d_x^2+\mu_n^2$, respectively.
The product over Matsubara sectors with $n>0$ 
represents quantum corrections to the prefactor. 
This product diverges at large $n$. In the imaginary part of free energy (\ref{ImF}) the divergence is absorbed by the logarithmic mass renormalization in the leading exponent, see \cite{Pirvu:2024nbe} for details. The renormalized product over $n>0$ in (\ref{factor}) is finite and in the classical (high-temperature) limit it goes to 1, so only the contribution of the $0^{\rm th}$ Matsubara sector remains.

It is convenient to compute the determinant of the operator (\ref{Omu}) at arbitrary $\mu$. Its ratio to the free determinant with the same $\mu$ is evaluated as \cite{Coleman:1978ae}:
\be \label{detform}
    \frac{\D_\mu}{\D_\mu^{(0)}} = \lim_{x\to+\infty}\frac{\vf_\mu(x)}{\vf_\mu^{(0)}(x)}\;.
\ee
Here $\vf_\mu$, $\vf_\mu^{(0)}$ are solutions of the equations
\be
\label{ModeEq1}
    {\cal O}(\mu)\vf_\mu(x) = 0 \;, \qquad {\cal O}^{(0)}(\mu)\vf_\mu^{(0)} = 0 \;,
\ee
with vanishing asymptotics at negative infinity:
\be 
    \vf_\mu = \vf_\mu^{(0)} = \e^{\mu x}  \quad \text{at}\quad x\to-\infty \;.
\ee
Clearly, $\vf_\mu^{(0)}= \e^{\mu x}$ at all $x$, while $\vf_\mu$ takes the form
\begin{align} \label{solphimu}
    \vf_\mu(x) &= c\, {\rm P}_2^\mu(\tanh x)\;, \\
    c &= -\mu(1-\mu)(2-\mu)\Gamma(-2-\mu)\;, \nonumber
\end{align}
where ${\rm P}_2^\mu$ is the associated Legendre polynomial. The asymptotics of (\ref{solphimu}) at positive infinity reads
\be \label{solphias}
    \vf_\mu(x)= c \, \frac{\e^{\mu x}}{\Gamma(1-\mu)} , \quad x \to +\infty\;.
\ee
Substituting it into Eq.~\eqref{detform}, we obtain
\be \label{fracdet}
    \frac{\D_\mu}{\D_\mu^{(0)}}=\frac{(\mu-m)(\mu-2m)}{(\mu+m)(\mu+2m)} \;.
\ee

The determinant in the $n=0$ sector vanishes because the bubble has a translational zero mode
\be \label{zeromode}
    \vf_{0,1}(x)=\sqrt{\frac{3\l T}{4m^3}}\, \phi_b^{\prime}(x) = - \sqrt{\frac{3Tm}{2}}\cdot\frac{\sh{mx}}{\ch^2{mx}}\;.
\ee
The coefficient is fixed by the normalization condition,
\be \label{zeronorm}
    \int \diff \tau \, \diff x\, \vf_{0,1}^2=1\;.
\ee
The zero mode satisfies Eq.~\eqref{eigen2} with $\mu_0=m$ and $\alpha_{0,1}=0$. To regularize the determinant, we take $\mu$ slightly different from $m$. The corresponding operator has a small eigenvalue $\a_{0,1}\approx \mu^2-m^2$. We divide out its contribution and obtain
\be \label{zerofactor2}
    \frac{\D_0}{\D^{(0)}_0}\mapsto 
    \frac{\D^{\prime}_0}{\D^{(0)}_0}=\lim_{\mu\to m}
    \frac{2\pi}{\mu^2-m^2}\frac{\D_\mu}{\D_{\mu}^{(0)}}=-\frac{2\pi}{12m^2}\;.
\ee
Note that we include a $2\pi$ factor because each mode in the Gaussian integration brings $\sqrt{2\pi/\alpha_I}$ with $\alpha_I$ being the corresponding eigenvalue.

The integral over the zero mode is replaced by the integral over the positions of the bubble. To this end, we introduce a unity into the path integral (\ref{Zgen}):
\be \label{shiftunity}
    \begin{split}
        1 = \int \diff b\, 
        & \left|\int \diff\tau \diff x\, \phi(\tau,x) \vf_{0,1}^{\prime}(x+b)\right|\\
        &\times\delta\left(\int \diff\tau \diff x\, \phi(\tau,x)\vf_{0,1}(x+b)\right) \;,
    \end{split}
\ee
and take the integration over $\diff b$ outside. The inner integral then runs over configurations orthogonal to ${\vf_{0,1}(x+b)}$. The saddle point of this integral is given by the shifted bubble $\phi_b(x+b)$. Due to the translation invariance, the inner path integral is the same for all $b$, hence the outer integral over $b$ just gives the total length $L$. In addition, we obtain a factor 
\be \label{prefzero}
    \left|\int \diff\tau \diff x\, \phi_b(x) \vf_{0,1}^{\prime}(x)\right| = \sqrt{\frac{4m^3}{3\l T}}\;.
\ee   
Note that the expression inside the square root coincides with the bubble action $S[\phi_b]$, as it should be \cite{Coleman:1978ae}. Combining Eqs.~\eqref{zerofactor2},~\eqref{prefzero} into the free energy 
\eqref{ImF} and substituting into Eq.~(\ref{GammaAffleck}) we obtain 
Eq.~\eqref{G_E2}.

\section{Initial state preparation}
\label{app:hmc}

In this Appendix we compare two methods of the initial state preparation: the non-perturbative HMC method \cite{Neal:2011mrf} used in this work and the perturbative Gaussian approximation \cite{Pirvu:2024nbe}. 

{\bf\emph{Hamiltonian Monte Carlo}} --- This method produces the initial state as a result of a sequence of iterations. We decompose the field and its canonical momentum $\pi\equiv\dot\phi$ at $t=0$ in Fourier modes,
\be \label{Fourier}
    \phi_i=\frac{1}{\sqrt{L}}\sum_{j=0}^{N-1}\e^{ik_j x_i}\tilde\phi_j \, ,
    \quad N\equiv\frac{L}{a}\, ,\quad k_j\equiv\frac{2\pi j}{L}\, ,
\ee
with $\tilde\phi_j^* = \tilde\phi_{N-j}$, and similarly for $\pi_i$.
Each HMC iteration contains two steps. The first step is the momentum update. During this step the momenta $\tilde{\pi}_j$ of all modes are drawn independently from the Gaussian distribution with the variance given by
\be 
\label{FourierPiVar}
    \langle |\tilde\pi_j|^2\rangle = T \; ,
\ee
while the coordinates $\tilde{\phi_j}$ are not changed. The second step is to evolve the system according to its equation of motion. We choose the time of the evolution $mt=1$, which allows the system to move sufficiently far from its previous state. 
Repeating the procedure many times we obtain the state drown from the Boltzmann ensemble with the probability $\propto \exp(-H/T)$, where $H$ is the full Hamiltonian of the system.

We perform 1000 HMC iterations for each state in the initial ensemble and use the resulting state as an initial condition for the simulation. 
Sometimes the false vacuum decay can happen during the HMC procedure. In this case we discard the HMC chain and start a new one.

{\bf\emph{Gaussian method}} --- After a rescaling of the fields and coordinates one can show \cite{Pirvu:2024nbe} that the interactions in the theory are proportional to a single dimensionless parameter $\lambda T/m^3$. Thus, at low temperature the field is weakly coupled and its state can be approximated by a collection of Gaussian fluctuations around the false vacuum with the Rayleigh-Jeans spectrum. This suggests to prepare the initial state by randomly drawing the complex amplitudes in the Fourier decomposition (\ref{Fourier}) 
from independent Gaussian distributions with the variances (\ref{FourierPiVar}) and
\be \label{FourierVar}
    \langle |\tilde\phi_j|^2\rangle = T/\Omega_{j}^{2} \; ,
\ee
where $\Omega_j$ are the lattice mode frequencies.

At $\lambda T/m^3\sim 0.1$ considered in our work, the effects of interactions are non-negligible. In particular, we find that using the bare field mass in $\Omega_j^2$ produces slow pulsations of the field spectrum, whereas it must be stationary in thermal equilibrium. The pulsations are eliminated by taking into account the thermal correction to the mass \cite{Boyanovsky:2003tc} which shifts the squares of lattice frequencies,
\be\label{mTh}
    \Omega_j^2 = \frac{2}{a^2}(1-\cos{ak_j}) + m_{th}^2 \; , \quad m_{th}^2 = m^2 - \frac{3\lambda T}{2m} \; . 
\ee
At low $k$ the shift is $\sim 15\%$ and cannot be neglected.

\begin{figure}[t]
	\center{	
		\begin{minipage}[h]{0.99\linewidth}
			\center{\includegraphics[width=1.0\linewidth]{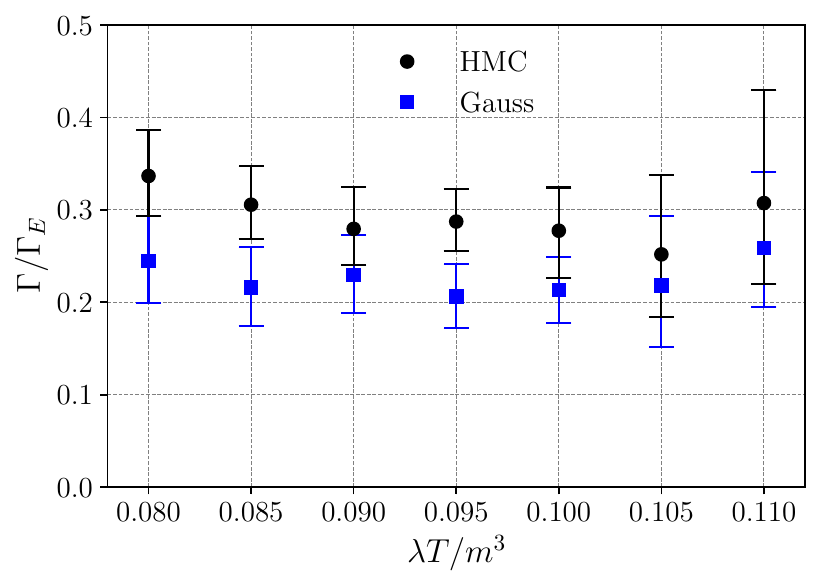}}
		\end{minipage}
	}
  \caption{The decay rate measured in simulations with the initial conditions prepared using the Gaussian state (\textit{blue squares}) and the HMC (\textit{black dots}).} 
	\label{fig:rate2}
\end{figure}

The false vacuum decay rate obtained with the Gaussian initial conditions is shown in 
Fig.~\ref{fig:rate2}. We see that it is systematically lower than the rate obtained with the HMC prepared state. A possible interpretation is that the Gaussian method underpopulates highly excited configuration of the field for which the non-linearities are enhanced. Though rare, these configurations are precursors to the critical bubble and reproducing their correct abundance is important for the accurate determination of the vacuum decay rate.

\section{Thermalization time}
\label{app:therm}

We can estimate the thermalization time by considering the Boltzmann equation for particle phase-space density $f_p$, see e.g.~\cite{Mueller:2002gd}.
In $(1+1)$ dimensions the leading processes resulting in the energy exchange between the particles are the $2\leftrightarrow 4$ and $3\leftrightarrow 3$ scatterings which give comparable contributions into the collision integral. For concreteness, let us focus on the former. Denoting the momenta of incoming particles by $p_1$, $p_2$, we have
\be
    \begin{split}
        \frac{\d f_{p_1}}{\d t}&  \simeq  \frac{1}{2\o_{p_1}}\int\frac{\diff \vec{p}_2\diff\vec{p}_3\diff \vec{p}_4\diff \vec{p}_5\diff \vec{p}_6}{(2\pi)^5 2\o_{p_2}2\o_{p_3}2\o_{p_4}2\o_{p_5}2\o_{p_6}} \\
        & \times (2\pi)^2\delta^{(2)}(p_1+p_2-p_3-p_4-p_5-p_6) |\mathcal{A}_{2\to 4}|^2 \\
        & \times \Bigl[ - f_{p_1}f_{p_2}(1+f_{p_3})(1+f_{p_4})(1+f_{p_5})(1+f_{p_6}) \\
        &\qquad + (1+f_{p_1})(1+f_{p_2})f_{p_3}f_{p_4}f_{p_5}f_{p_6}  \Bigr] \;.
    \end{split}
\ee
Assume for simplicity that all particles have comparable momenta of order $p$. Then the scattering amplitude is $\mathcal{A}_{2\to 4}\sim \l^2/\o_p^2$,
and the dominant contribution from the Bose-enhancement factor is $f_p^5 \sim (T/\o_p)^5$. This yields, 
\be \label{cross-section}
    \frac{1}{f_p}\frac{\d f_p}{\d t}\sim \frac{1}{(2\pi)^3}\frac{\l^4T^4}{\o_p^{11}}\;,
\ee
whence we read off the thermalization time  
\be \label{t_therm2}
    t_{th} \sim \frac{(2\pi)^3}{m} \left(\frac{m^3}{\l T} \right)^4\left(\frac{\o_p }{m} \right)^{11} \, .
\ee
The modes relevant for decay have frequencies $\omega_p\sim m$ and thermalize on the time scale (\ref{t_therm}). Note that the thermalization time grows steeply with the increasing frequency.  

Repeating the same argument in $(3+1)$-dimensional theory, where $2\leftrightarrow2$ processes are relevant, we obtain (see also \cite{Aarts:1996qi})
\be
\label{t_therm3}
t_{th}\sim \frac{(2\pi)^2}{m}\left(\frac{m}{\l T} \right)^2\left(\frac{\o_p }{m} \right)\;.
\ee
In this case the thermalization time scales as the inverse square of the temperature.

\section{Finding the unbiased decay rate}
\label{app:extrap}

Here we describe the numerical procedure used to find the unbiased decay rate in the cases when the dissipation is slow, $t_{th}>(\Gamma t)^{-1}$, and the survival probability is plagued by the classical Zeno effect discussed in Sec.~\ref{sec:Psurv}.

\begin{figure}[t]
	\center{	
		\begin{minipage}[h]{0.99\linewidth}
			\center{\includegraphics[width=1.0\linewidth]{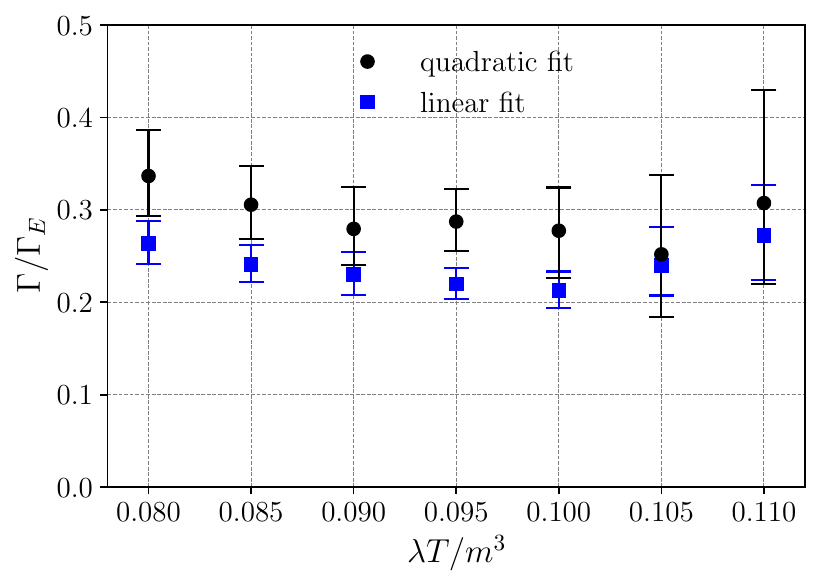}}
		\end{minipage}
	}
  \caption{The unbiased decay rate measured in simulations using the extrapolation procedure with the linear fit (\textit{blue squares}) and the quadratic fit (\textit{black dots}). The initial conditions are prepared with the HMC.
}
	\label{fig:rate3}
\end{figure}

We restrict the analysis to the part of the survival probability curve $y=\ln P_{surv}(t)$ with $y>0.9$ where the time-dependence introduced by the Zeno effect is expected to be small. We also remove the time interval $t<50$ where the transient effects are strong. The remaining part of the curve is split into 8 segments $(t_1,t_2),(t_2,t_3),...$ such that $y_2-y_1=y_3-y_2=...$. Each segment is fit by a straight line with the slope $\Gamma_i$, $i=1,..,8$. The error of the fit is estimated as $\sigma_{\Gamma_i}=\Gamma_i/\sqrt{N_i}$ where $N_i$ is the number of decay events within the $i$'th segment. It assumes that the survival probability follows the Gaussian distribution when $N_i\gg 1$, which we verified numerically using the bootstrap method.

We take $\ln \Gamma_i$ to be the logarithm of the decay rate $\Gamma(t)$ at $t=(t_i+t_{i+1})/2$. We then fit $\ln \Gamma(t)$ by the quadratic function
$\ln \Gamma(t) = a+bt+ct^2$ and take 
its intercept $a$ as the unbiased decay rate $\Gamma(T)$, to be compared with the theoretical prediction (\ref{G_E2}).
This extrapolation procedure is different from the one used in Ref.~\cite{Pirvu:2024nbe} where the fitting function was restricted to the linear form. We have found that this restriction introduces a systematic bias in the results, underestimating the rate and its errors, as shown in Fig.~\ref{fig:rate3}. We have checked with the bootstrap method that the error estimated using the quadratic fit is correct.

\bibliography{Refs.bib}

\end{document}